\documentclass[twocolumn,showpacs,preprintnumbers,amsmath,amssymb]{revtex4}
\usepackage{amssymb}
\usepackage[dvips]{graphicx}
\begin{document}

\title{Normal state Nernst effect, semiconducting-like resistivity and diamagnetism of  underdoped  cuprates}
\author{A. S. Alexandrov}

\affiliation{Department of Physics, Loughborough University, Loughborough, United Kingdom\\
}

\begin{abstract}
Semiconducting-like low-temperature in-plane resistivity indicates
that there are no remnants of superconductivity above the resistive
phase transition at $T > T_c$ in underdoped cuprates. The model with
the chemical potential pinned near the mobility edge inside the
charge-transfer optical gap describes quantitatively
 the  Nernst effect, thermopower, diamagnetism and the
unusual low-temperature  resistivity of underdoped cuprates as
normal state phenomena above $T_c$.
\end{abstract}

\pacs{74.40.+k, 72.15.Jf, 74.72.-h, 74.25.Fy} I
\maketitle

In the framework of the weak-coupling BCS theory the superconducting
state is described  by a nonzero Gor'kov anomalous average
 $\cal{F}(\mathbf{r,r^{\prime
}})=\langle
\psi_{\downarrow}(\mathbf{r})\psi_{\uparrow}(\mathbf{r}^{\prime
})\rangle $, which is zero above the resistive phase transition
temperature $T_c$. When the BCS theory is extended to the
strong-coupling regime, electrons are paired into lattice
bipolarons, which are real-space  pairs dressed by phonons,
\emph{both} below and above $T_c$ \cite{alebook}.  The state above
$T_{c}$ is a normal charged   Bose-liquid and  below $T_{c}$ phase
coherence  of the preformed bosons sets in. In this regime
$\cal{F}(\mathbf{r,r^{\prime }})$  describes bosons in the
Bose-Einstein condensate similar to the Bogoliubov anomalous average
of the annihilation operator in the Bose-gas. As in the BCS theory
the state above $T_{c}$ is perfectly "normal" in the sense that the
off-diagonal order parameter $\cal{F}(\mathbf{r,r^\prime })$ is zero
at $T>T_c$.

In disagreement with the weak-coupling BCS  and the strong-coupling
 bipolaron theories a significant fraction of research in the field
of  superconducting cuprates claims that the superconducting
transition is only a phase ordering while the superconducting order
parameter $\cal{F}(\mathbf{r,r^{\prime }})$ remains nonzero above
the resistive $T_c$. One of the key experiments supporting this
viewpoint is the large Nernst signal observed  in the normal  state
of cuprates  (see \cite{xu,cap,cap2} and references therein). Refs
\cite{xu,ong}  propose a "vortex scenario", where the long-range
phase coherence is destroyed by mobile vortices, but the amplitude
of the off-diagonal order parameter remains finite and the Cooper
pairing with a large binding energy exists  well above $T_c$
supporting the so-called  "preformed Cooper-pairs" or "the phase
fluctuation" model \cite{kiv}. The model is based on the assumption
that superfluid density is small compared with  the normal carrier
density in cuprates.  These claims seriously undermine many
theoretical and experimental  works on superconducting cuprates,
which consider the state above $T_c$ as perfectly normal with no
off-diagonal order.

However, the vortex  scenario is unreconcilable with the extremely
sharp resistive transitions at $T_c$ in high-quality samples of
cuprates. For example, the in-plane and out-of-plane resistivity of
$Bi-2212$, where the anomalous Nernst signal has been measured
\cite{xu}, is perfectly normal above $T_c$, showing only a few
percent positive or negative magnetoresistance \cite{zavale}.  The
preformed Cooper-pairs model \cite{kiv}  is clearly incompatible
with a great number of thermodynamic, magnetic,  kinetic and optical
measurements, which show that only holes (density $x$), doped into a
parent insulator are carriers \emph{both} in  the normal and the
superconducting states of cuprates. The assumption \cite{kiv} that
the superfluid density $x$ is small compared with the normal-state
carrier density $1-x$ is also inconsistent with the theorem
\cite{leg}, which proves that the number of supercarriers at $T=0$K
should be the same as the number of normal-state carriers in any
clean superfluid.

Faced with these inconsistences  we have recently described the
unusual Nernst signal in overdoped $La_{1.8}$Sr$_{0.2}$CuO$_4$  in a
different manner as the normal state phenomenon \cite{alezav}. Here
we  extend our description to cuprates with low doping level
accounting  not only for  their anomalous Nernst signal, but also
for the thermopower, normal state diamagnetism and a
semiconducting-like in-plane low-temperature resistivity as observed
in recent \cite{xu,cap,cap2,ong} and more earlier experiments.

In underdoped cuprates strong  on-site repulsive correlations
(Hubbard $U$) are essential in shaping the insulating state of
parent compounds. The  Mott-Hubbard insulator arises from a
potentially metallic half-filled band as a result of the Coulomb
blockade of electron tunnelling to neighboring sites \cite{mott}.
The first band to be doped in cuprates is the oxygen band inside the
Hubbard gap. The strong electron-phonon interaction (see for
experimental facts Ref. \cite{alebook} ) creates  oxygen hole
polarons and  inter-site bipolarons. Hence the chemical potential
remains inside the optical charge-transfer gap, as clearly observed
in the tunnelling experiments by Bozovic et al. \cite{boz0}.
Disorder, inevitable with doping, creates localised impurity states
for holes separated by a mobility edge from their extended states
like in conventional amorphous semiconductors \cite{mott,ell}. Then
the chemical potential should
 be found at or near the mobility edge in slightly doped cuprates,
if they superconduct.

Naturally carriers, localised below the mobility edge, contribute to
the normal-state longitudinal transport together with the itinerant
carriers in extended states. On the other hand, the contribution of
localised carriers of any statistics to the \emph{ transverse}
transport is usually small as in many amorphous semiconductors
\cite{ell}. Importantly, if the localised-carrier contribution is
not negligible, it \emph{adds} to the contribution of itinerant
carriers to produce a large Nernst signal, $e_{y}(T,B)\equiv
-E_{y}/\nabla _{x}T$, while it \emph{reduces} the thermopower $S$
and the Hall angle $\Theta$. This unusual "symmetry breaking" is  at
variance with ordinary metals where the familiar "Sondheimer"
cancelation \cite{sond} makes $e_{y}$ much smaller than $S\tan
\Theta$ because of the electron-hole symmetry near the Fermi level.
Such  behavior originates in the "sign" (or "$p-n$") anomaly of the
Hall conductivity of localised carriers. The sign of their Hall
effect is often $opposite$ to that of the thermopower as observed in
many amorphous semiconductors \cite{ell} and described theoretically
\cite{fri}.

The Nernst signal can be expressed in terms of the kinetic
coefficients $\sigma _{ij}$ and $\alpha _{ij}$ as
\begin{equation}
e_{y}={\frac{{\sigma _{xx}\alpha _{yx}-\sigma _{yx}\alpha
_{xx}}}{{\sigma _{xx}^{2}+\sigma _{xy}^{2}}}},
\end{equation}
where the current density  is given by $j_{i}=\sigma
_{ij}E_{j}+\alpha _{ij}\nabla _{j}T$.
 When the chemical potential $\mu$ is at the mobility edge,
  the localised carriers contribute to the transport,
 so  $\sigma _{ij}$ and $\alpha _{ij}$ in Eq.(1) can be expressed
as $\sigma^{ext} _{ij}+\sigma^{l}_{ij}$ and $\alpha^{ext}
_{ij}+\alpha^{l}{ij}$, respectively \cite{alezav}. Since the Hall
mobility of carriers localised below $\mu$, $\sigma^{l}_{yx}$, has
the  sign opposite to that of carries in the extended states above
$\mu$, $\sigma^{ext}_{yx}$, the sign of the off-diagonal Peltier
conductivity $\alpha^{l}_{yx}$ should be the same as the sign of
$\alpha^{ext}_{yx}$. Then  neglecting the magneto-orbital effects in
the resistivity (since $\Theta \ll 1$ \cite{xu}) we obtain
\begin{equation}
S\tan \Theta \equiv {\sigma _{yx}\alpha _{xx}\over{\sigma
_{xx}^{2}+\sigma _{xy}^{2}}} \approx\rho (\alpha ^{ext}_{xx}-|\alpha
^{l}_{xx}|) (\Theta^{ext}-|\Theta^{l}|)
\end{equation}
and
\begin{equation}
e_{y}\approx\rho (\alpha^{ext} _{yx}+|\alpha^{l} _{yx}|)-S\tan
\Theta,
\end{equation}
where $\Theta^{ext}\equiv \sigma^{ext}_{yx}/\sigma_{xx}$,
$\Theta^{l}\equiv \sigma^{l}_{yx}/\sigma_{xx}$, and
$\rho=1/\sigma_{xx}$ is the resistivity.

\begin{figure}
\begin{center}
\includegraphics[angle=-90,width=0.50\textwidth]{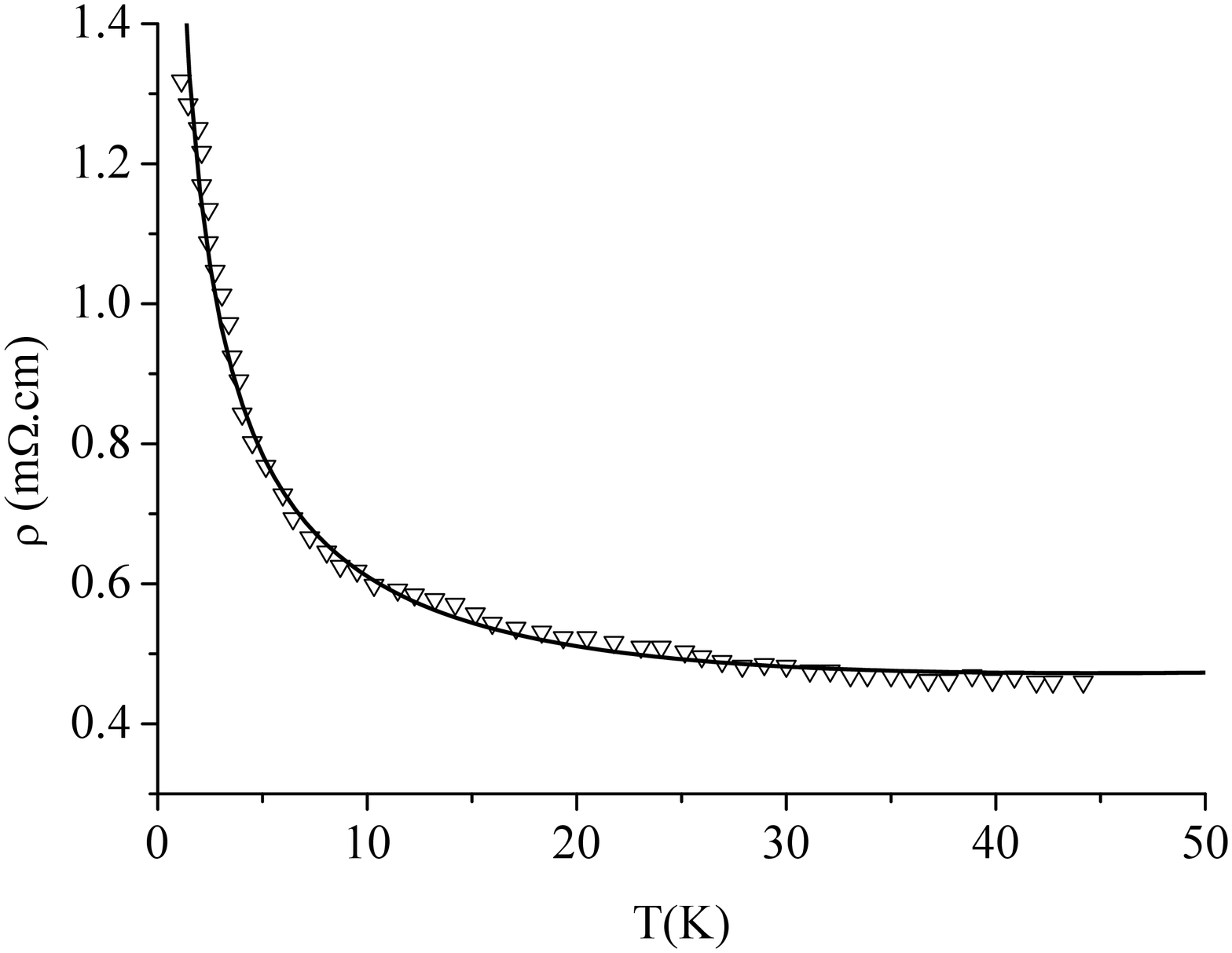}
\vskip -0.5mm \caption{Normal state in-plane resistivity of
underdoped La$_{1.94}$ Sr$_{0.06}$CuO$_4$ (triangles \cite{cap}) as
revealed in the field $B=12$ Tesla  and compared with the bipolaron
theory, Eq.(6) (solid line).}
\end{center}
\end{figure}
Clearly the model, Eqs.(2,3) can account for a low value of
$S\tan\Theta$ compared with a large value of $e_y$ in underdoped
cuprates \cite{xu,cap2} because of the sign anomaly. Even in the
case when localised carriers contribute little to the conductivity
their contribution to the thermopower, $S=\rho (\alpha
^{ext}_{xx}-|\alpha ^{l}_{xx}|))$, could almost cancel  the opposite
sign contribution of itinerant carriers. Indeed, if the carriers are
bosons, their longitudinal conductivity in two-dimensions,
$\sigma^{ext} \propto \int_0 dE E df(E)/dE$ diverges logarithmically
when $\mu$ in the Bose-Einstein distribution function
$f(E)=[\exp((E-\mu)/T)-1]^{-1}$ goes to zero and the relaxation time
$\tau$ is a constant (here and further we take $\hbar=c=k_B=1$). At
the same time $\alpha^{ext}_{xx}\propto \int_0 dE E(E-\mu) df(E)/dE$
remains finite, and it could have a magnitude comparable   with
$\alpha^{l}_{xx}$. Statistics of bipolarons  effectively changes
from Bose to Fermi-like statistics with lowering energy below the
mobility edge because of the Coulomb repulsion of bosons in
localised states \cite{alegile}. Hence one can use the same
expansion near the mobility edge as in  ordinary amorphous
semiconductors to obtain the familiar textbook result $S=S_0T$ with
a constant $S_0$ at low temperatures \cite{mott3}.

The model becomes particularly simple, if we   neglect the localised
carrier contribution to $\rho$, $\Theta$ and $\alpha_{xy}$, and take
into account that $\alpha^{ext}_{xy} \propto B/\rho^2$ and
$\Theta^{ext}\propto B/\rho$ in the Boltzmann theory. Then Eqs.(2,3)
yield
\begin{equation}
S\tan \Theta  \propto T/\rho
\end{equation}
and
\begin{equation}
e_{y}(T,B)\propto (1-T/T_1)/\rho.
\end{equation}

According to our earlier suggestion \cite{alelog} the
semiconducting-like  dependence of $\rho(T)$ in underdoped cuprates
(\cite{cap,cap2} and references therein) at low temperatures
originates from the elastic scattering of non-degenerate itinerant
carriers by charged impurities, different from  scenarios based on
any kind of metal-insulator transitions. The relaxation time of
\emph{non-degenerate} carriers  depends on temperature as $\tau
\propto T^{-1/2}$ for scattering by short-range deep potential
wells, and as $T^{1/2}$ for scattering by very shallow wells as
discussed in Ref. \cite{alelog}. Combining both scattering rates
yields
\begin{equation}
\rho =\rho_0[(T/T_2)^{1/2}+(T_2/T)^{1/2}].
\end{equation}
Eq.(6) with $\rho_0=0.236$ m$\Omega\cdot$cm and $T_2=44.6$K fits
 well the experimental semiconducting-like normal state
resistivity of underdoped La$_{1.94}$ Sr$_{0.06}$CuO$_4$ in the
whole low-temperature range from  2K up to 50K, Fig.1,  as revealed
in the field $B=12$ Tesla \cite{cap,cap2}. Another high quality fit
can be  obtained combining the Brooks-Herring formula for the 3D
scattering off screened charged impurities, as proposed in
Ref.\cite{kast} for almost undoped $LSCO$, or the Coulomb scattering
in 2D ($\tau \propto T$) and a temperature independent scattering
rate off neutral impurities with the carrier exchange \cite{erg}
similar to the scattering of slow electrons by hydrogen atoms in
three dimensions. Hence the scale $T_2$, which determines the
crossover toward an insulating behavior, depends on the relative
strength of two scattering mechanisms. Importantly the expressions
(4,5) for  $S\tan \Theta$ and $e_y$ do not depend on particular
scattering mechanisms, but only on the experimental $\rho(T)$.
Taking into account the excellent fit of Eq.(6) to the experiment,
these expressions can be parameterized as
\begin{equation} S\tan \Theta = e_0
{(T/T_2)^{3/2}\over{1+T/T_2}},
\end{equation}
and
\begin{equation}
e_{y}(T,B)=e_0{(T_1-T) (T/T_2)^{1/2}\over{T_2+T}} ,
\end{equation}
where $T_1$ and $e_0$ are temperature independent.
\begin{figure}
\begin{center}
\includegraphics[angle=270,width=0.50\textwidth]{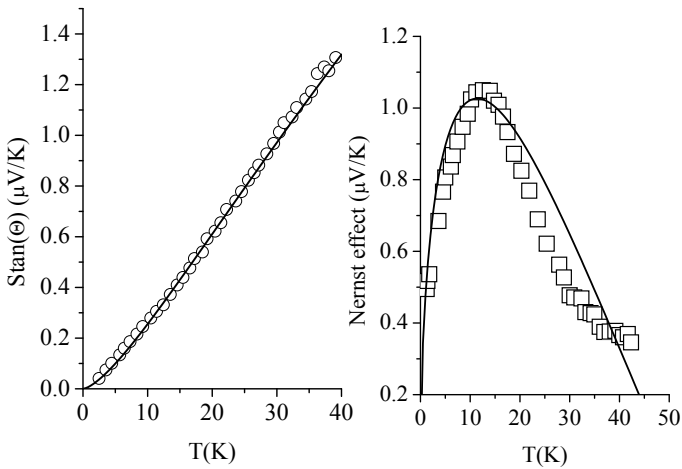}
\vskip -0.5mm \caption{$S\tan\Theta$ (circles \cite{cap2} )  and the
Nernst effect $e_y$  (squares \cite{cap})  of underdoped La$_{1.94}$
Sr$_{0.06}$CuO$_4$ at $B=12$ Tesla compared with the bipolaron
theory, Eqs.(7,8) (solid lines).}
\end{center}
\end{figure}

In spite of many simplifications, the model describes  remarkably
well both $S\tan \Theta$ and $e_y$  measured in La$_{1.94}$
Sr$_{0.06}$CuO$_4$ with a $single$ fitting parameter, $T_1=50$K
using the experimental $\rho(T)$. The constant  $e_0=2.95$ $\mu$V/K
scales the magnitudes of $S\tan \Theta$ and $e_y$.  The magnetic
field $B=12$ Tesla destroys the superconducting state of the
low-doped La$_{1.94}$ Sr$_{0.06}$CuO$_4$ down to $2$K, Fig.1, so any
residual superconducting order above $2$K is clearly ruled out. At
the same time   the Nernst signal, Fig.2, is remarkably large. The
coexistence of the large Nernst signal and a nonmetallic resistivity
is in sharp disagreement with the vortex scenario, but is in
agreement with our model. Taking into account the field dependence
of the conductivity of localised carriers, their contribution to the
transverse magnetotransport and the phonon-drug effect (at elevated
temperatures) can well describe the magnetic field dependence of the
Nernst signal \cite{alezav} and improve the fit in Fig.2 but at the
expense of the increasing number of fitting parameters.

Another experimental observation, which has been  linked with the
Nernst signal and mobile vortexes above $T_c$ \cite{ong}, is
enhanced diamagnetism. A number of experiments (see, for example,
\cite{mac,jun,hof,nau,igu,ong} and references therein), including
torque magnetometries, showed enhanced diamagnetism near and above
$T_c$, which has been explained as  fluctuation diamagnetism in
quasi-2D superconducting cuprates (see, for example Ref.
\cite{hof}). The data taken at relatively low magnetic fields
(typically below 5 Tesla) revealed a crossing point in the
magnetization $M(T,B)$ of most anisotropic cuprates (e.g.
$Bi-2212$), or in $M(T,B)/B^{1/2}$ of less anisotropic $YBCO$
\cite{jun}. The dependence of magnetization (or $M/B^{1/2}$) on the
magnetic field has been shown to vanish at some characteristic
temperature below $T_c$ in agreement with  conventional
fluctuations. However the data taken in high magnetic fields (up to
30 Tesla) have shown that the crossing point, anticipated for
low-dimensional superconductors and associated with superconducting
fluctuations, does not explicitly exist in magnetic fields above 5
Tesla \cite{nau}. Most surprisingly the torque magnetometery
\cite{mac,nau} uncovered a diamagnetic signal somewhat above $T_c$
which \emph{increases} in magnitude with applied magnetic field.

Here we argue that such behaviors are incompatible with the vortex
scenario but can be understood with bipolarons. Accepting the vortex
scenario and fitting the magnetization data for $Bi-2212$ with the
conventional logarithmic field dependence \cite{ong}, one obtains
surprisingly high upper critical fields $H_{c2} > 120$ Tesla  even
at temperatures close to $T_c$, and a very large Ginzburg-Landau
parameter, $\kappa=\lambda/\xi
>450$ . The in-plane
low-temperature magnetic field penetration depth is $\lambda\approx
200$ nm in optimally doped $Bi-2212$ (see, for example \cite{tal}).
Hence the zero temperature coherence length $\xi$ turns out to be
about the lattice constant, $\xi=0.45$nm, or even smaller. Such a
small coherence length rules out the "preformed Cooper pairs"
\cite{kiv}, since the pairs are virtually not overlapped at any size
of the Fermi surface in $Bi-2212$. Moreover the magnetic field
dependence of $M(T,B)$ at and above $T_c$ is entirely inconsistent
with what one expects of a vortex liquid. While $-M(B)$  decreases
logarithmically at temperatures well below $T_c$, the  experimental
curves \cite{mac,nau,ong} clearly show that $-M(B)$  increases with
the field at and  above $T_c$ , just the opposite of what one could
expect in a vortex liquid.  This significant departure from the
London liquid behavior clearly indicates that the vortex liquid does
not appear above the resistive phase transition \cite{mac}.

Some time ago \cite{den} we  proposed  that
  anomalous diamagnetism $M(T,B)$ in cuprates could be the Landau
normal-state diamagnetism of preformed bosons.  When the strong
magnetic field is applied perpendicular to the copper-oxygen plains
the quasi-2D bipolaron energy spectrum is quantized,
$E=\omega(n+1/2) +2t_c [1-\cos(k_zd)]$, where $\omega=2eB/m_b$,
$n=0,1,2,...$, and $t_c$, $k_z$, $d$ are the hopping integral, the
momentum and the lattice period perpendicular to the planes.
Differentiating the thermodynamic potential  one can readily obtain
$M(0,B)= -n_b \mu_b$ at low temperatures, $T\ll T_c$, which is the
familiar Schafroth's result \cite{sha}. Here $n_b$ is the bipolaron
density, $\mu_b=e/m_b$ is the "bipolaron" Bohr magneton, and $m_b$
is the bipolaron in-plane mass. The magnetization of charged bosons
is field-independent at low temperatures. At high temperatures, $T
\gtrsim T_c$  the bipolaron gas is almost classical. The
experimental conditions are such that $T \gg \omega$, when $T$ is of
the order of $T_c$ or higher, so $M(T,B)\approx -n_b \mu_b
\omega/6T$.
 It is the familiar Landau  orbital diamagnetism  of non-degenerate carriers.
The  bipolaron in-plane mass in cuprates is about $m_b\approx 10
m_e$  as follows from a number of independent experiments and
numerical (QMC) simulations \cite{alebook}. Using this mass yields
$M(0,B) \approx 2000$ A/m with the bipolaron density $n_b=10 ^{21}$
cm$^{-3}$. Then the magnitude and the field/temperature dependence
of $M(T,B)$ at and above $T_c$ are about the same as experimentally
observed in Refs \cite{nau,ong}.

The pseudogap temperature $T^*$, which is half of the bipolaron
binding energy in the model, depends on the magnetic field  because
of  spin-splitting of the single-polaron band by the magnetic-field.
Also the singlet-triplet exchange energy of inter-site bipolarons
depends on the field for the same reason. As a result the number of
singlet bipolarons and thermally excited triplet pairs and single
polarons
 depend on the field and on the temperature. When the depletion of the bipolaron density
with temperature and magnetic field
 is taken into account, the crossing point in $M(T,B)$
disappears at high magnetic fields as observed, and the normal state
 magnetization of singlet bipolarons fits
experimental $M(T,B)$ curves \cite{ong}  in the whole normal state
and critical regions \cite{alemag}.

In summary, we have described the normal state Nernst effect, the
thermopower, the diamagnetism and the semiconducting-like in-plane
resistivity of underdoped cuprates at low temperatures as the
normal-state properties of non-degenerate oxygen holes doped into
the Mott-Hubbard charge-transfer insulator with the chemical
potential close to the mobility edge. The familiar "sign" (or
"$p-n$") anomaly of the Hall conductivity of localised carriers
accounts for a small value of $S\tan\Theta_H$ compared with a large
value of $e_y$. The semiconducting-like temperature dependence of
the in-plane resistivity at low temperatures originates from the
elastic scattering of non-degenerate itinerant carriers by charged
impurities, rather than from any localisation. The enhanced
diamagnetism at $T > T_c$ is the normal state  orbital diamagnetism
of bipolarons.

  I thank  V.V. Kabanov and V.N. Zavaritsky for
valuable discussions. The work was supported by  EPSRC (UK) (grant
EP/C518365/1).

\end{document}